\documentclass[runningheads]{workshop}
\usepackage[english]{babel}

\usepackage{graphicx}  % standard LaTeX graphics tool
\usepackage{caption}
\begin{document}
 
\title*{Numerical approaches for investigating the chaotic behavior of multidimensional Hamiltonian systems}
 
\titretab{Numerical approaches for investigating the chaotic behavior of multidimensional Hamiltonian systems
 % \protect\newline uncomment if your title is too long
 }
 
\titrecourt{Investigating the chaotic behavior of multidimensional Hamiltonian systems}

\author{Haris Skokos\inst{1}}

\index{Skokos Haris}

\auteurcourt{H. Skokos}
 
\adresse{$^1$ Nonlinear Dynamics and Chaos Group, Department of Mathematics and Applied Mathematics, University of Cape Town, Rondebosch, 7701, South Africa}
\email{haris.skokos@gmail.com}

\maketitle              

%\begin{center}
%\bf Session  % Uncomment one of the lines below
%History of chaos
%Data analysis
%Nonlinear dynamical systems theory
%Environmental dynamics (hydrology, epidemiology, oceanography, climatology)
%Applications
%\end{center}
%\vspace{0.4cm}

%================================
\begin{abstract}
	We discuss various numerical approaches for studying the chaotic dynamics of multidimensional Hamiltonian systems, focusing our analysis on the chaotic evolution of initially localized energy excitations in the disordered Klein-Gordon oscillator chain in one spatial dimension. 
\end{abstract}

%================================
\section{Introduction}
\label{sec:intro}

Disordered Hamiltonian systems with many degrees of freedom serve as valuable models to mimic the natural heterogeneity found in the real world. Typically in such systems, random values are assigned to one of their parameters for each degree of freedom. Disordered systems find practical applications in describing various important physical processes, such as material conductivity, light propagation in optical waveguides, Bose-Einstein condensates dynamics, granular solids' structural behavior, and the behavior of DNA molecules, while at the same time are ideal test beds for studying the dynamical properties of multidimensional Hamiltonian models.

One of the foundational discoveries in the field of disordered systems is the phenomenon known as `Anderson localization', first introduced by Anderson in 1958 \cite{A58}, which  refers to the spatial confinement of energy excitations and the half of spreading in a linear lattice, in the presence of strong enough disorder. However, the introduction of nonlinearity into such systems has sparked significant interest in recent years, both in theory, simulations and experiments (see e.g.~\cite{PS08,REFFFZMMI08,JKA10,LDTRZMLDIM11,LIF14,B14}). Numerical investigations particularly concerning one-dimensional (1D) Hamiltonian lattice models like the nonlinear disordered Klein-Gordon (DKG) oscillator chain, have shown that nonlinear interactions  result to the destruction of  energy localization, leading to the eventual subdiffusive spreading of wave packets in two different dynamical regimes, the so-called `weak' and `strong chaos' spreading regimes \cite{FKS09,SKKF09,LBKSF10}.

It is now well-established that energy spreading in disordered lattices is a chaotic process. Here we  perform a brief overview of various numerical techniques which have been successfully implemented for revealing the specific characteristics of this chaotic behavior. We  use the DKG model to present these methods, although these techniques can be applied for studying the chaotic behavior of any multidimensional Hamiltonian system.

%================================
\section{Numerical investigation of the chaotic behavior of the one-dimensional nonlinear disordered Klein-Gordon lattice}
\label{sec:DKG}

The Hamiltonian function of the 1D DKG lattice system of $N$ anharmonic oscillators is
\begin{equation}
	H = \sum_{i = 1}^{N} \left[ \frac{p_i ^2}{2} + \frac{\tilde{\epsilon}_l }{2} q_i ^2 +
	\frac{q_i ^4}{4} + \frac{1}{2W}\left( q_{i+1} - q_{i} \right) ^2 \right],
	\label{eq:H_DKG}
\end{equation}
where $q _i$ and $p_i$ are respectively  the generalized position and momentum of site $i$, $\tilde{\epsilon}_l$ are disorder parameters of the on-site potential whose values are  uniformly chosen from  the interval $\left[\frac{1}{2}, \frac{3}{2} \right]$ (a particular set of $\epsilon_i$, is referred to as a disorder realization of the system), and $W>0$ determines the disorder strength. The Hamiltonian function (\ref{eq:H_DKG}) is an integral of motion and its value (usually referred to as the system's energy $E$) remains constant in time and also serves as a control parameter of the model's nonlinearity strength.
All numerical results presented here are obtained by following the evolution of energy distributions (wave packets) created by the initial excitation of $L$  central oscillators at the same energy level, $E/L$,  by setting $p_i=\pm \sqrt{2E/L}$, with randomly assigned signs (for $L=1$ we use the $+$ sign) for the excited $L$ sites, and $q_i=0$ for all sites. We consider  fixed boundary conditions, i.e.~$q_0=p_0=q_{N+1}=p_{N+1}=0$, taking care  that the studied lattice is large enough so that the energy does not reach its boundaries until the end of the integration.  We define  normalized energy distributions $\xi _i = \left[\frac{p_i ^2}{2} + \frac{\tilde{\epsilon}_i }{2} q_i^2 + \frac{q_i^4}{4} +
\frac{1}{4W} \left( q_{i + 1} - q_{i} \right)^2\right]/E$, and compute their second moment $m_2 = \sum _i (i - \bar{i})^2 \xi_l$,  which measures the distribution's extent,  along with the participation number $P = 1/\sum _i \xi _i ^2$, which estimates the number of highly excited sites, with $\bar{i} = \sum _i i \xi _i$ being the position of the distribution's center. In \cite{FKS09,SKKF09,LBKSF10,SMS18} the existence of two spreading dynamical regimes was theoretically predicted and numerically verified, namely the so-called weak and strong chaos regimes, which are characterized by particular power law increases of the wave packet's $m_2$ and $P$. More specifically, $m_2$  increases in time $t$ as $m_2(t) \propto t^{a_m}$ with $a_m=1/3$ ($a_m=1/2$) for the weak (strong) chaos case, while  $P$ grows as $P(t) \propto t^{a_p}$ with $a_p=1/6$ ($a_p=1/4$). The generality of these findings is supported by the fact that these exponential growths were also observed for the 1D disordered discrete nonlinear Schr\"{o}dinger (DDNLS) equation \cite{PS08,FKS09,SKKF09,LBKSF10,SMS18}.

\begin{figure}
	\centering
	\includegraphics[width=0.64\columnwidth,keepaspectratio]{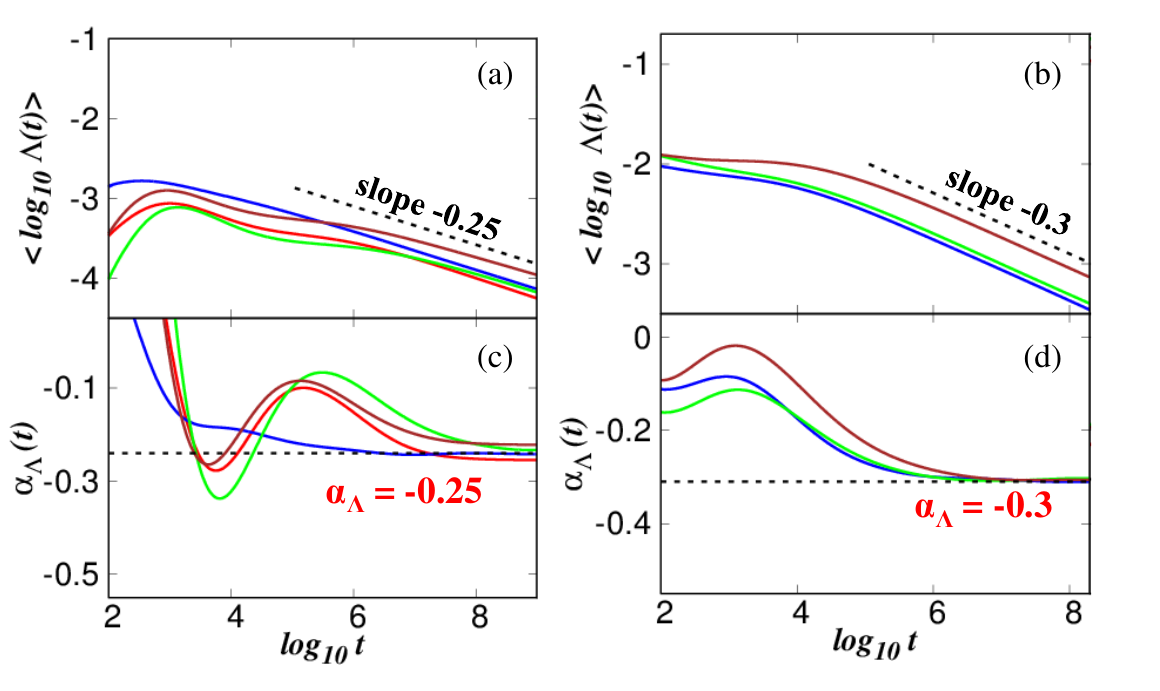}
	\caption{Averaged (and smoothed) results over 100 disorder realizations of the time evolution of [(a), (b)] the ftmLE $\Lambda$ (\ref{eq:ftMLE}),  and [(c), (d)] the corresponding numerically computed derivatives $\alpha_\Lambda$ 
		for different [(a), (c)] weak chaos and [(b), (d)] strong chaos cases. The straight dashed lines indicate slopes [(a), (c)] $\alpha_\Lambda =-0.25$ and [(b), (d)]  $\alpha_\Lambda =-0.3$. The presented  weak chaos cases in (a) and (c) correspond to the following initial energy densities $\xi_i$ and parameter sets  $W$, $L$: $\xi_i = 0.01$, $W=3$, $L=37$ (red curves); $\xi_i = 0.4$, $W=4$, $L=1$ (blue curves); $\xi_i = 0.01$, $W=4$, $L=21$ (green curves); $\xi_i = 0.02$, $W=5$, $L=13$ (brown curves). The strong chaos cases in (b) and (d) correspond to $\xi_i = 0.1$, $W=2$, $L=83$ (blue curves); $\xi_i = 0.1$, $W=3$, $L=37$ (green curves); $\xi_i = 0.1$, $W=3$, $L=83$, (brown curves). (After \cite{SMS18}) }
	\label{fig:1}
\end{figure}

The system's chaoticity can be quantified by its maximum Lyapunov  exponent (mLE) $\Lambda_1$ (see e.g.~\cite{S10} and references therein), which is estimated as the limit for $t\rightarrow \infty$ of the finite-time mLE (ftmLE)
\begin{equation}
	\Lambda (t) = \frac{1}{t}\ln \frac{\| \vec{w}(t)
		\|}{\| \vec{w}(0) \|},
	\label{eq:ftMLE}
\end{equation}
i.e.~$\Lambda_1 = \lim_{t\to\infty} \Lambda(t)$, with $\vec{w}(t) = \left( \delta q_1 (t), \ldots, \delta q_N (t), \delta p_1 (t), \ldots, \delta p_N (t) \right)$ being a phase space deviation vector from the studied orbit  at time $t$, and  $ \| \cdot \|$ denoting the usual Euclidian norm of a vector. We note that the deviation vector's evolution is governed by the so-called `variational equations' (see e.g.~\cite{S10} for more details). In the case of regular motion,  $\Lambda(t)$ tends to zero following the power law $\Lambda(t) \propto t^{-1}$, 
while for chaotic orbits, it tends to a non zero positive value. In \cite{SGF13,SMS18} it was found that the ftmLE of initially localized excitations of the 1D DKG system 
exhibits a power law decay of the form  $\Lambda(t) \propto t^{\alpha_{\Lambda}}$ with $\alpha_{\Lambda}$ being different from the $\alpha_{\Lambda}=-1$ value observed in cases of regular motion. In particular,  $\alpha_{\Lambda}\approx -0.25$ ($\alpha_{\Lambda}\approx -0.3$) for the weak  (strong) chaos case (see Fig.~\ref{fig:1}). This behavior indicates that, as the number of excited lattice sites increases due to the wave packet spreading, the system becomes less chaotic, although the  dynamics does not  show any indication of crossover to regular behavior, at least up to the computationally  accessible times). The same $\alpha_{\Lambda}$ values were also found for the 1D DDNLS system \cite{SMS18}.  
%%%%%%%%%%%%%%%%%%%%%%%%%%%%%%%%%%%%
%%%%%%%%%%%%%%%%%%%%%%%%%%%%%%%%%%%%

\begin{figure}
	\centering
	\includegraphics[scale=0.27]{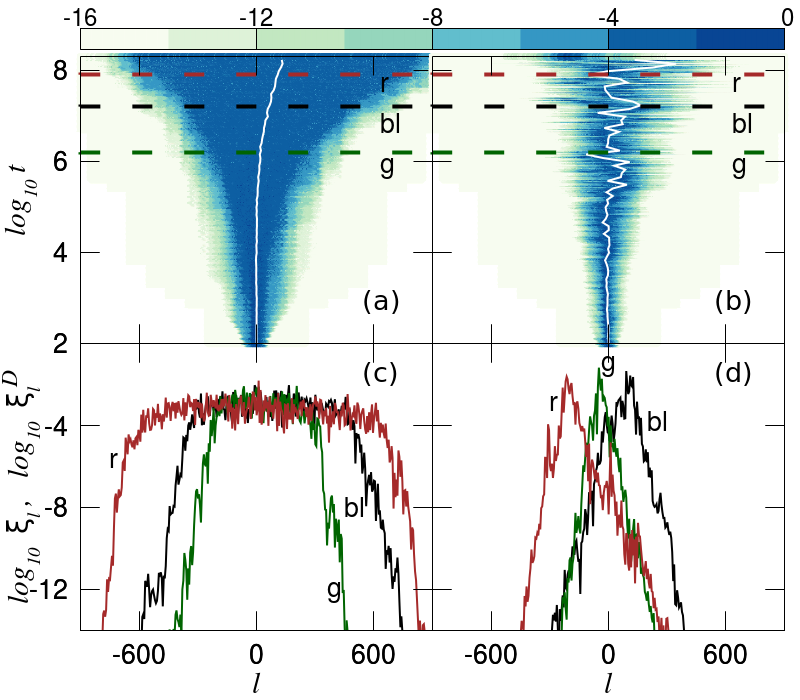}
	\caption{The dynamics of a representative initial condition of the strong chaos case with $\xi_i = 0.1$, $W=3$, $L=83$, for one disorder realization.  Time evolution of (a) the normalized energy distribution  and (b) the corresponding DVD. The color scales at the top of the figure are used for coloring lattice sites according to their (a) $\log_{10} \xi_i$  and (b) $\log_{10} \xi_i^D$  values. In both panels a white curve traces the distribution's center. Normalized energy distributions (c) and  DVDs  (d)  at times $\log_{10}t=6.2$ [green curves (g)], $\log_{10}t=7.2$ [black curves (bl)] and  $\log_{10}t=7.9$ [red curves (r)]. These times are also denoted by similarly colored horizontal dashed lines in (a) and (b). Note that the lattice site index is denoted by $l$. (After \cite{SMS18})}
	\label{fig:2}
\end{figure}

In order to better understand  the mechanisms of the chaotic wave packet spreading in the weak and strong chaos regimes, and to identify the sites which  behave more chaotically at any given time,  the deviation vector distribution (DVD)   
\begin{equation}
	\label{eq:DVD}
	\xi^D_i(t) = \frac{\delta q_i(t) ^2 + \delta p_i(t)^2}{\sum_i \left[ \delta q_i(t)^2 + \delta p_i(t)^2 \right]}, \,\,\, i=1,2,\ldots, N,
\end{equation}
created by the time evolution of the  vector $\vec{w} (t)$ used for the computation of $\Lambda$  (\ref{eq:ftMLE}), was introduced and applied for the study of the 1D DKG system in \cite{SGF13},  and later on  was also successfully  implemented  for both the 1D DKG and DDNLS models \cite{SMS18}. The main idea behind the introduction of the DVD is that since $\vec{w} (t)$ eventually aligns to the most unstable direction in the system's phase space (which corresponds to the mLE), large $\xi^D_i$ values will indicate at which lattice sites the sensitive dependence on initial conditions is higher. Thus, the DVD can be used to visualize the motion of chaotic seeds inside the spreading wave packet. In \cite{SGF13,SMS18} it was found that, for both the weak and strong chaos regimes, the DVD remains always inside the excited part of the lattice, retaining a rather localized, pointy shape, indicating that at any given time only a small fraction of the excited sites are highly chaotic. Nevertheless, these chaotic hot spots do not always contain  the same oscillators but they meander through the system, supporting the homogeneity of chaos inside the wave packet. A representative strong chaos case is shown in Fig.~\ref{fig:2}. From the results of this figure  we see that the energy distribution [Figs.~\ref{fig:2}(a) and (c)] expands  to  larger regions of the lattice in a, more or less, symmetric fashion around the position of the initial excitation as the evolution of the distribution's mean position [white curve in Fig.~\ref{fig:2}(a)] is rather smooth,   remaining close to the lattice's center. On the other hand, the DVD [Figs.~\ref{fig:2}(b) and (d)] does not actually spread and at first is  located in the region of the initial excitation,  but it starts moving around widely after $\log_{10}t \approx 6$, something which is clearly depicted in the  evolution of its mean position [white curve in Fig.~\ref{fig:2}(d)], showing random fluctuations with increasing amplitude. 
%%%%%%%%%%%%%%%%%%%%%%%%%%%%%%%%%%%%
%%%%%%%%%%%%%%%%%%%%%%%%%%%%%%%%%%%%

Despite the usefulness of the mLE as a chaos indicator, this quantity is an average measure of chaoticity, providing information about the global behavior of a dynamical system. As such, the value of the mLE by itself is not enough to reveal the characteristics of the spatiotemporal evolution of active chaotic regions in a multidimensional system. Thus, the implementation of the so-called `frequency map analysis' (FMA) technique \cite{L90} was used in \cite{SGF22} to follow in time the chaotic or regular behavior of each individual lattice site. The main idea of the FMA is that the evaluation of the  fundamental frequency  of an observable produced by the evolution of the coordinates $q_i(t)$ and $p_i(t)$, $i=1,2, \ldots, N$, of each oscillator $i$ can be used to identify chaotic behavior. In the case of regular motion, i.e.~motion on an torus, the fundamental frequency of the produced time series will not change in time  due to the quasi-periodic nature of the underlying phase space orbit. So, the computed fundamental frequencies $f_{1i}$ and $f_{1i}$ in two successive time windows should practically coincide, something which is not expected for chaotic orbits. So, the relative change of these two frequencies, quantified by
\begin{equation}
	D_i=\left| \frac{f_{2i}-f_{1i}}{f_{1i}} \right|, \, \, \, i=1,2, \ldots, N,
	\label{eq:D}
\end{equation}
can be used to identify the chaotic or regular nature of motion, because small $D_i$ values denote the practical constancy of the fundamental frequency and consequently indicate regular motion, while large $D_i$ values signify chaotic behavior characterized by strong variations in the computed frequency values. 

\begin{figure}
	\centering
	\includegraphics[width=0.94\columnwidth,keepaspectratio]{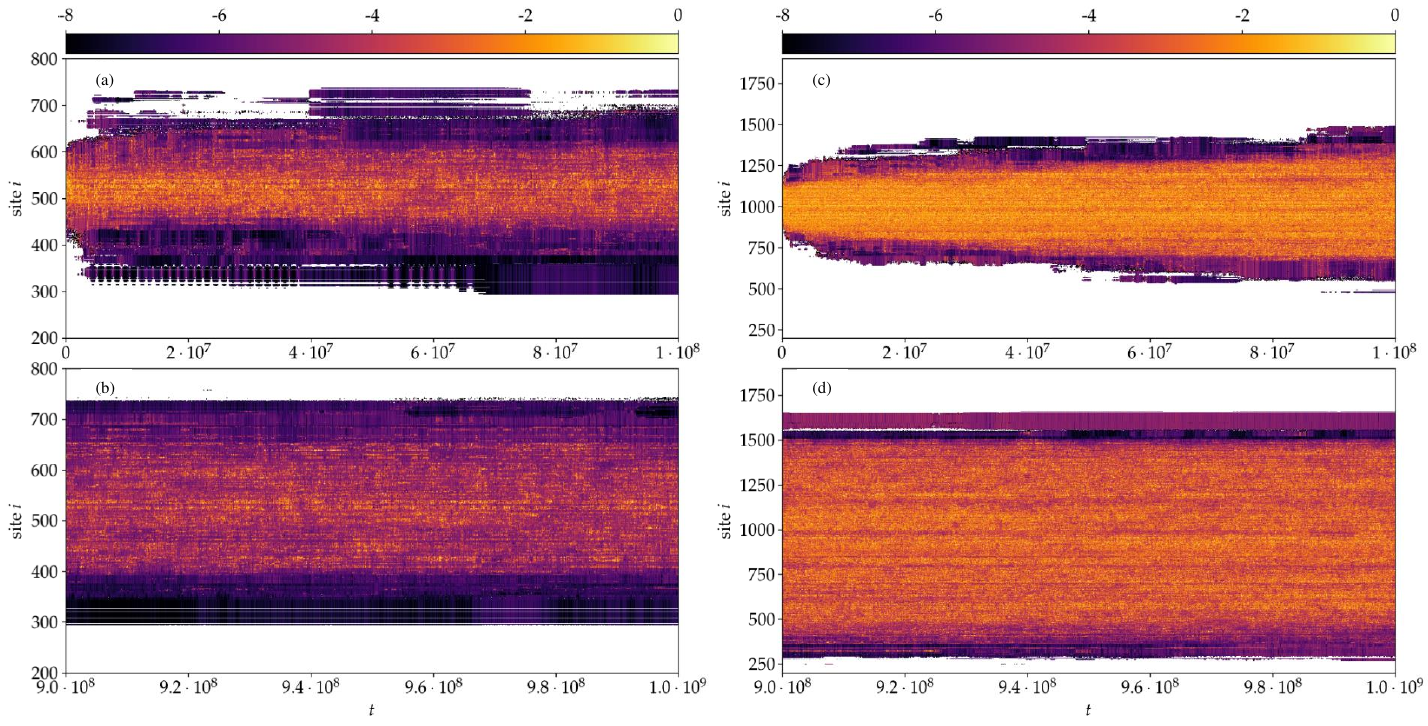}
	\caption{The time evolution of the quantity $D_i$ (\ref{eq:D}) for [(a), (b)] a weak chaos case with $\xi_i = 0.4$, $W=4$, $L=1$, and [(c), (d)] a strong chaos case with $\xi_i = 0.2$, $W=4$, $L=21$ in the time intervals [(a), (c)] $0 \leq t \leq 10^8$ and [(b), (d)] $9 \cdot 10^8 \leq t \leq 10^9$. In each panel the horizontal axis is time $t$ in linear scale, while the vertical axis shows the  site number $i$. The color scales at the top of the upper row of panels are used for coloring lattice sites according to their $\log_{10} D_i$ values.	(After \cite{SGF22})}
	\label{fig:3}
\end{figure}

The implementation of the FMA in \cite{SGF22} facilitated the  visualization  of chaos evolution in the propagated energy distribution, revealed several characteristics of the dynamics related to the location of highly chaotic oscillators, and permitted the identifications of  differences between the weak and strong chaos  regimes. More specifically, it was shown that chaotic behavior appears at the central regions of the wave packet, where the energy density is relatively large, and that the  fraction of highly chaotic oscillators decreases in time. With respect to the differences between the weak and strong chaos regimes it was found that the chaotic component of the wave packet is more extended in the strong chaos case, while at the same time the fraction of sites behaving chaotically is much higher in the strong chaos regime, typically around 5 times higher than in the weak chaos one. All these conclusions can also be drawn from the results of Fig.~\ref{fig:3} where the wave packet evolution of a representative weak chaos [Figs.~\ref{fig:3}(a) and (b)] and strong chaos [Figs.~\ref{fig:3}(c) and (d)] case is shown for  two time windows of length $10^8$ time units: immediately after the initial excitation [Figs.~\ref{fig:3}(a) and (c)] and after $9 \cdot 10^8$ time units [Figs.~\ref{fig:3}(b) and (d)]. We note that each site  is colored according to its $\log_{10} D_i$ value.
%%%%%%%%%%%%%%%%%%%%%%%%%%%%%%%%%%%%
%%%%%%%%%%%%%%%%%%%%%%%%%%%%%%%%%%%%

\begin{figure}[ht]
	\centering 
	\includegraphics[width=0.53\columnwidth,keepaspectratio]{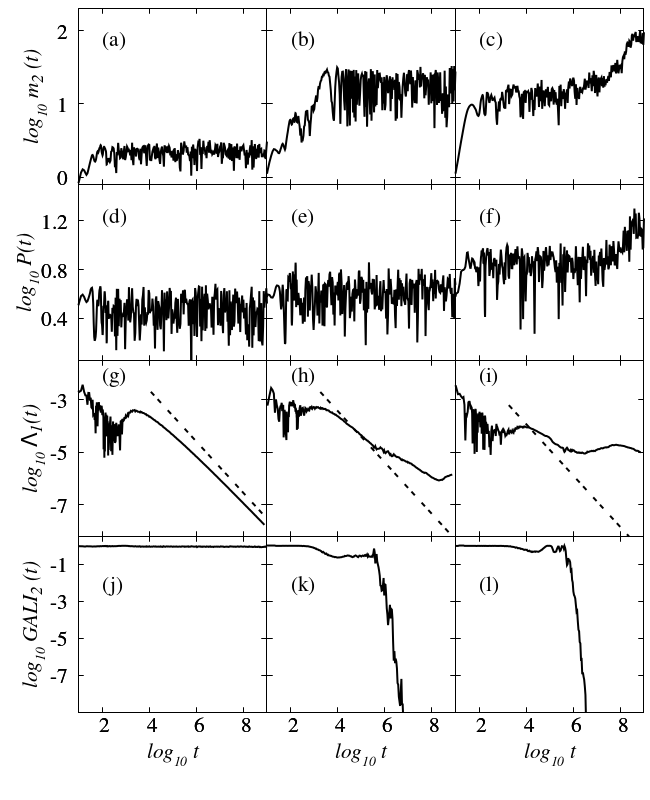}
	\caption{The time evolution of [(a)-(c)] $m_2$, [(d)-(f)] $P$, [(g)-(i)] $\Lambda$ (\ref{eq:ftMLE}) (which is denoted as $\Lambda_1$) and [(j)-(l)] GALI$_2$ (\ref{eq:gali})  for the same single site ($L=1$) excitation of the DKG system (\ref{eq:H_DKG}) with $W=6$, $\xi_i=0.02$, for three different disorder realizations (one per column). The disorder realization of the left column leads to a regular evolution, while the one used in the middle  and in the right column, respectively correspond to localized and spreading chaos. [(g)-(i)] The straight dashed lines guide the eye for slope $-1$, which corresponds to regular dynamics. (After \cite{SS22})}
	\label{fig:4}
\end{figure}

The Generalized Alignment Index of order $k$ (GALI$_k$) \cite{SBA07} is an efficient chaos detection technique whose main advantage over the computation of the ftmLE (\ref{eq:ftMLE}) is its ability to identify chaos much more clearly and efficiently \cite{SM16}. The index is computed through the evolution of $k$ initially orthonormal deviation vectors $\vec{w}_i$, and  its value is computed as the norm of the wedge product of these normalized vectors $\hat{\vec{w}}_i$
\begin{equation}
	\label{eq:gali}
	\mbox{GALI}_k(t)=||\hat{\vec{w}}_1(t)\wedge\hat{\vec{w}}_2(t)\wedge\cdots\wedge\hat{\vec{w}}_k(t)||.
\end{equation} 
In \cite{SS22} the GALI$_2$ method (which is equivalent to the so-called Smaller Alignment Index (SALI) \cite{S01}) was successfully implemented for discriminating between localized and spreading chaos in the 1D DKG model. In Fig.~\ref{fig:4} representative cases of three different dynamical behaviors, namely  regular dynamics (left column), localized (middle column) and spreading chaos (right column) are presented for single site excitations of different disorder lattices, but for the same initial condition, total energy and disorder strength. We see that GALI$_2$  decreases exponentially fast to zero in the case of chaotic orbits [Figs.~\ref{fig:4}(k) and (i)], while it remains practically constant for regular orbits [Fig.~\ref{fig:4}(j)].
%%%%%%%%%%%%%%%%%%%%%%%%%%%%%%%%%%%%
%%%%%%%%%%%%%%%%%%%%%%%%%%%%%%%%%%%% 

Based on extensive simulations like the ones of Fig.~\ref{fig:4}  the probabilistic nature  of the appearance of chaotic or regular behaviors when the system's nonlinearity decreases (leading the DKG model closer to its linear limit), was clearly shown in \cite{SS22}. Furthermore, it was found that  below some small, but not negligible, energy threshold all initial conditions lead to regular motion. In addition, the existence  of a higher energy threshold above which all considered initial conditions and system arrangements lead to a chaotic wave-packet spreading was also identified. For energies between these two thresholds the GALI$_2$ method managed to efficiently determined the percentages of spreading and localized chaos.

%\vfill

%================================
{}

\end{document}